# Coherent space-gated microscopy: a step towards deep-tissue phase imaging of biological cells

Mooseok Jang[1,2†*], Hakseok Ko[1,2†], Won Kyu Lee[3], Jae-Seung Lee[3], and Wonshik Choi[1,2*]

[1]Center for Molecular Spectroscopy and Dynamics, Institute for Basic Science (IBS), 145 Anam-ro, Seongbuk-gu, Seoul 02841, Korea.
[2]Department of Physics, Korea University, 145 Anam-ro, Seongbuk-gu, Seoul 02841, Korea.
[3]Department of Materials Science and Engineering, Korea University, 145 Anam-ro, Seongbuk-gu, Seoul 02841, Korea

*Correspondence to: majng@korea.ac.kr (M.J.), wonshik@korea.ac.kr (W.C.)
†These authors contributed equally to this work.

High-resolution optical microscopy suffers from a low contrast in scattering media where a multiply scattered wave obscures a ballistic wave used for image formation. To extend the imaging depth, various gating operations – confocal, coherence, and polarization gating – have been devised to filter out the multiply scattered wave. However, these gating methods are imperfect as they all act on the detection plane located outside a scattering medium. Here, we present a new gating scheme, called 'space' gating, that rejects the multiply scattered wave directly at the object plane inside a scattering medium. Specifically, we introduced a 30 µm-wide acoustic focus to the object plane and reconstructed a coherent image only with the ballistic wave modulated by acousto-optic interaction. This method allows us to reject the multiply scattered wave that the existing gating methods cannot filter out and improves the ratio of the ballistic wave to the multiply scattered wave by more than 100 times for a scattering medium more than 20 times thicker than its scattering mean free path. Using the coherent imaging technique based on space gating, we demonstrate the unprecedented imaging capability - phase imaging of optically transparent biological cells fully embedded within a scattering medium - with a spatial resolution of 1.5 µm.

Improving the imaging depth of high-resolution optical microscopy has been a long-standing goal in the field of bioimaging due to its impact on biological studies and optical diagnostics[1]. For an ideal diffraction-limited imaging, the main strategy is to detect the so-called 'ballistic' wave that propagates straight through a scattering medium and carries an intact object information. However, the ballistic wave is quickly obscured by the multiply scattered wave even at a shallow depth as its intensity decays exponentially due to multiple light scattering. To extend the imaging depth, the prevailing approach so far has been to filter out the multiply scattered wave by applying various gating operations such as confocal[2,3], time (or coherence)[4–7], and polarization gating[8,9]. For example, optical coherence tomography - one of the most successful biomedical imaging modalities – greatly extends the imaging depth by combining all these gating operations[7,10–12]. Similarly, spatial correlation within a time-gated transmission or reflection matrix has recently been used to selectively extract an image information[13,14]. Furthermore, various adaptive optics approaches have been proposed to maintain the effectiveness of gating operations even in the presence of sample-induced aberration[15–18].

Despite of all these tremendous advances, the imaging depth has not yet reached the detection limit set by the dynamic range of state-of-the-art sensor technology. The ballistic wave is, in principle, detectable even for the depths greater than 15 $l_s$ in an epi-detection geometry (where $l_s$ is the scattering mean free path of the scattering medium) if an image sensor with a high dynamic range (e.g. 1:10$^4$) is used in conjunction with the interferometric detection converting an intensity recording to an electric field measurement[10,13,14,19]. Currently, the imaging depth limit is instead set by the competition between a ballistic wave and a multiply scattered wave that bypasses the existing gating operations. The residual multiply scattered wave can be significantly stronger than the ballistic wave well before reaching the detection limit[10,19]. For instance, the chance that a multiply scattered wave has the similar flight time as a ballistic wave and passes through the time gating of finite width increases with the imaging depth. Likewise, a large fraction of multiply scattered wave can pass through a confocal pinhole in an extreme turbidity, and in turn mistakenly considered as a ballistic wave. In fact, these imperfections of existing gating methods are mainly due to their action at the detection plane located outside a scattering medium. To reach the detection limit, it is critical to develop an additional gating method whose mechanism is independent of the existing methods and yet effective enough to complement them.

Here we propose a new gating scheme, called 'space' gating, based on the acousto-optic modulation of light[20,21]. We have realized the concept of space gating by selectively measuring the ballistic wave that is modulated by the high-frequency ultrasound focus whose size is as small as ~ 30 μm × 70 μm. Unlike confocal or time gating, space gating is directly applied on the object plane inside a scattering medium for rejecting the multiply scattered wave whose optical path spreads beyond the extent of ultrasound focus. Therefore, it removes the multiply scattered wave that cannot be filtered out by the existing gating operations. Furthermore, the modulated ballistic wave is measured by the means of either interferometric confocal detection[22] or coherent aperture synthesis[23,24]. This enables coherent imaging – phase imaging as well as the amplitude imaging – in scattering media which would be a critical advantage for deep-tissue label-free imaging of transparent biological cells. Considering that the previous acousto-optic imaging techniques have used all the modulated wave as a whole[25,26] and that their spatial resolutions have been typically dictated by the size of ultrasound focus, the proposed concept of space gating represents the first acousto-optic approach that selectively uses only the ballistic wave for coherent imaging with an ideal optical diffraction-limited resolution.

With the aid of space gating, we have demonstrated imaging of amplitude objects through the scattering layers thicker than $23l_s$ with the optical diffraction-limited resolution of 1.5 μm. In this regime, the intensity ratio between the ballistic and multiply scattered waves, which was initially lower than 1/10 only with confocal gating, was improved to above 10 by the addition of space gating. Therefore, the ratio between the ballistic and multiply scattered waves was enhanced by more than 100 times. Furthermore, combining the noise rejection capability of space gating with the advantage of the coherent treatment of ballistic wave, we have demonstrated quantitative phase imaging of biological cells fully embedded within a scattering medium. Phase imaging is a powerful tool of label-free imaging as it enables visualizing the difference in refractive indices or thicknesses of transparent objects such as biological cells[27,28]. However, phase imaging in a scattering medium has not yet been realized because the phase retardation is much more susceptible to multiple scattering than the amplitude variation of the probing beam. The proposed concept of space gating, as an independent and complementary addition to the existing gating operations, sets an important step towards reaching the fundamental depth limit of diffraction-limited imaging relying on the ballistic wave and also opens a new avenue for label-free imaging of biological cells through scattering tissues.

**Principle**

Figure 1a illustrates the concept of space gating combined with confocal gating. In order to understand the effect of space gating, we first describe the source of noise in confocal gating in terms of an illumination transfer function $T_i(r_o; r_i)$ and a detection transfer function $T_d(r_o; r_d)$.

Here $r_o$ is an arbitrary position on the object plane, $r_i$ is the illumination point of a focused laser beam, and $r_d$ is the detection point on the sensor plane that is optically conjugate to the object plane. $T_i(r_o; r_i)$ describes the illumination field on the object plane for a focused illumination at the position $r_i$, and $T_d(r_o; r_d)$ describes the electric field at the detection point $r_d$ for the point source departing from each object point $r_o$. Figure 1b1 and 1c1 show the typical transfer functions measured on the object plane for the specific illumination and detection points (see Methods for details of the measurement procedure). When the sample is transparent, $|T_i(r_o; r_i)|^2$ and $|T_d(r_o; r_d)|^2$ are simply the intensity maps created by the focused ballistic wave (Fig. 1b1). On the other hand, they appear as the focused spot superimposed with the randomly fluctuating speckle pattern in the presence of scattering (Fig. 1c1). Thus, we may decompose each transfer function into the ballistic ($S$) and multiply scattered components ($M_i$ or $M_d$):

$$T_i(r_o; r_i) = S(r_o; r_i)e^{-\frac{L_i}{2l_s}} + M_i(r_o; r_i) \qquad (1)$$

$$T_d(r_o; r_d) = S(r_o; r_d)e^{-\frac{L_d}{2l_s}} + M_d(r_o; r_d). \qquad (2)$$

Here, $L$ and $l_s$ are the thickness and scattering mean free path of the sample, and $S(r_o; r_i)$ and $S(r_o; r_d)$ denote the transfer functions of ballistic waves in the absence of scattering. And the subscript i and d indicate the illumination and detection part of the sample as indicated in Fig. 1a. The optical paths associated with $S$ is indicated by green lines, and $M_i$ and $M_d$ by blue lines (both solid and dotted) in Fig. 1a.

The transmitted field at the detection point $r_d$ for the focused illumination at the position $r_i$ is then given as the integration of the product of illumination and detection transfer functions with respect to $r_o$ over the entire object plane $R$:

$$E(r_d; r_i) = \int_R T_i(r_o; r_i) T_d(r_o; r_d) dr_o = E_S(r_d; r_i) + E_M(r_d; r_i). \tag{3}$$

Since our main interest is in biological imaging where forward scattering is dominant, we neglected backward scattering between the illumination and detection parts of the scattering medium. $T_i(r_o; r_i) T_d(r_o; r_d)$ in Eq. (3) can be considered as the contribution map describing how each point $r_o$ on the object plane contributes to the light propagation of the focused illumination at $r_i$ to detection point $r_d$. Using Eqs. (1) and (2), we obtain four terms from the multiplication – one term $[S(r_o; r_i) S(r_o; r_d) e^{-\frac{(L_i + L_d)}{2l_s}}]$ associated with the 'signal' field $E_S(r_o; r_i)$, and three terms $[S(r_o; r_i) e^{-\frac{L_i}{2l_s}} M_d(r_o; r_d), S(r_o; r_i) e^{-\frac{L_d}{2l_s}} M_i(r_o; r_d)$, and $M_i(r_o; r_i) M_d(r_o; r_d)]$ associated with the multiply scattered 'noise' field $E_M(r_o; r_i)$. The first two terms in $E_M(r_o; r_i)$ are typically negligible in a highly scattering regime because the ballistic wave is much weaker than the scattered wave in its total power (i.e. $S(r_o; r_i) e^{-\frac{L_i}{l_s}} \ll \int_R |M_i(r_o; r_i)|^2 dr_o$ and $S(r_o; r_d) e^{-\frac{L_d}{l_s}} \ll \int_R |M_d(r_o; r_d)|^2 dr_o$). Therefore, the noise field is approximately given as

$$E_M(r_d; r_i) \sim \int_R M_i(r_o; r_i) M_d(r_o; r_d) dr_o. \tag{4}$$

Figures 1b2 and 1c2 show the typical contribution maps $|T_i(r_o; r_i) T_d(r_o; r_d)|^2$ defined on the object plane when the detection point $r_d$ is conjugate to the illumination point $r_i$ (i.e. when confocal gating is used). In the contribution map (Fig. 1c2), the peak in the center is mostly contributes to the signal field $E_S(r_d; r_i)$ with the relatively weak contribution to the noise field $E_M(r_d; r_i)$. On the other hand, the object point $r_o$ outside the center peak solely contributes to the fluctuating noise field $E_M(r_d; r_i)$. In other words, the signal field $E_S(r_d; r_i)$ travels through the confined region of the peak on the object plane, while the noise field $E_M(r_d; r_i)$ propagates through the widely extended area as expressed in Eq. (4). Considering the detected field $E(r_d; r_i)$ at $r_d$ in this confocal system is the integration of the contribution map over the entire object plane (Eq. (3)), the integrated contribution to the multiply scattered noise can easily dominate the ballistic signal even when the ballistic wave appears as the peak in the contribution map.

The proposed concept of space gating is to set a spatial window $R_{SG}$ around $r_i$ (or $r_d$) on the object plane (as indicated as the red spot in Fig. 1a and the dashed circles in Fig. 1b2 and 1c2) in a way that only the wave transmitted through the gating window contributes to the detected field. In our experiments, we implemented this concept by modulating the optical wave with a high frequency acoustic focus and selectively detecting the modulated wave emerged from the acoustic focus. With the space gating, the detected field $E^{SG}(r_d; r_i)$ [$= E_S^{SG}(r_d; r_i) + E_M^{SG}(r_d; r_i)$] is determined by the integration of contribution map $T_i(r_o; r_i)T_d(r_o; r_d)$ only within $R_{SG}$ (i.e. the right side of Eq. (3) with $R_{SG}$ in place of $R$). Because the ballistic wave that appears as a peak on the contribution map lies within $R_{SG}$, the signal field is unaffected by the space gating, i.e. $E_S^{SG}(r_d; r_i) = E_S(r_d; r_i)$. In contrast, the noise field can be significantly suppressed as the space gating excludes all the noise contribution $M_i(r_o; r_i)M_d(r_o; r_d)$ from the object point $r_o$ located outside $R_{SG}$ (indicated as the dotted blue lines in Fig. 1a).

We quantify the effect of space gating by the noise suppression factor $\eta$, which is defined by the intensity ratio of multiply scattered wave at $r_d$ without and with space gating. In a highly scattering regime where the Eq. (4) is applied, $\eta$ can be estimated as the area ratio between the spatial extent of the intrinsic noise contribution $|M_i(r_o; r_i)M_d(r_o; r_d)|^2$ and $R_{SG}$:

$$\eta = |E_M(r_d; r_i)|^2 / |E_M^{SG}(r_d; r_i)|^2 \sim \min(\Delta w_{M_i}, \Delta w_{M_d})^2 / \Delta w_{SG}^2. \tag{5}$$

Here, we approximate the multiply scattered component of illumination and detection transfer functions $|M_i(r_o;r_i)|^2$ and $|M_d(r_o;r_d)|^2$ as the top-hat functions with their respective widths of $\Delta w_{M_i}$ and $\Delta w_{M_d}$. $\Delta w_{SG}$ is the width of $R_{SG}$ set by the acoustic focus size in our experiments. Therefore, the larger the spatial extent of the transfer functions is and the smaller the spatial window $R_{SG}$ is, the larger the effect of space gating is. $\Delta w_{M_i}$ and $\Delta w_{M_d}$ depend on many parameters such as the average number of scattering events $L/l_s$, the anisotropy constant $g$, and the geometry of the sample and the layout of the optical system. For biological tissues, $\Delta w_{M_i}$ and $\Delta w_{M_d}$ can be typically in the range from hundreds of microns to millimeters when $L/l_s \sim 10$, from which we can expect $\eta > 100$ if the size of space gating $\Delta w_{SG}$ is as small as tens of microns such as in a high-frequency acoustic focus.

**Results**

**Point spread functions of confocal imaging setup with space-gating**

Figure 2a shows the experimental configuration of the confocal imaging system integrated with an acousto-optic space gating. The illumination beam was focused on the object, and the transmitted beam was collected by an objective lens and then focused onto the detector plane. The numerical apertures (NA) of the objective lenses on the illumination and detection paths were 0.18, setting the diffraction-limit resolution to be 1.5 μm. Three cycles of the focused acoustic wave whose frequency was $f_{US} = 50$ MHz was temporally synchronized with 532-nm laser pulse of 7 ns width at the repetition rate of 40 kHz.

The scheme to implement the space gating is based on an interferometric detection similar to the previously demonstrated ultrasound-modulated optical tomography[25,26,29,30]. When light propagates through the oscillating pressure field at the acoustic focus, a fraction of the light is modulated by the frequency of $f_{US}$. The complex field of the modulated wave is then selectively measured using 4-step phase-shifting interferometry[31]. The frequency of the reference beam is set at $f_0 + f_{US}$ to form a static interference pattern with the modulated wave. The unmodulated wave having the frequency of $f_0$, albeit much larger than the modulated wave, remains as a constant background independent of phase stepping due to the out-of-phase relation with the reference beam.

Although the interferometric confocal detection provides the phase map of the ballistic wave, the phase drift during the focal scanning deteriorates the phase image of the object. Therefore, to achieve quantitative phase imaging, we switch the illumination beam to a plane wave and vary the incidence angle for coherent aperture synthesis where the coherent (i.e. both amplitude and phase) image is synthesized in a way that the ballistic wave is collectively accumulated [23,24]. Note that once the ballistic wave is properly accumulated for every incidence angle constituting the focused beam in confocal detection, the signal to noise ratio of coherent aperture synthesis is identical to that of confocal method. However, in most of our experiments where we imaged amplitude objects, we used the confocal scheme shown in Fig. 2a because the confocal scheme provides a higher signal to noise ratio for the initial detection of ballistic wave before reconstructing the image.

To confirm the spatial extent of the acousto-optic space gating, we illuminated the transparent sample with a plane wave. Figures 2b and c present the measured intensity map without and with space gating summed over 900 incidence angles. The reference beam and the focused acoustic beam were switched off for the conventional measurement without space gating. Without space gating, the intensity map was uniform across the field of view (Fig. 2b), indicating that every object location equally contributes to the measured signal. When the space gating was applied, only the wave traveling through the gating window $R_{SG}$, defined by the ultrasound focus, was visible (Fig. 2c). The full widths at half maximum (FWHMs) of the space gating were 29 μm and 72 μm along the *x*- and *y*-axis, respectively. The gating contrast, measured by the ratio between the average intensity inside the blue box and outside the orange box in Fig. 2c, was about 100, and the modulation efficiency around the focal area (see Methods for calculation details) was 5%. Note that this acousto-optic modulation efficiency does not affect the signal $\left|E_S^{SG}(r_\mathrm{d};r_\mathrm{i})\right|^2$ to noise $\left|E_M^{SG}(r_\mathrm{d};r_\mathrm{i})\right|^2$ ratio and the noise suppression factor $\eta$ because both the signal and noise are subject to the same modulation efficiency.

Figures 2d and 2e present the point spread functions (PSF) $|E(r_\mathrm{d};r_\mathrm{i})|^2$ and $|E_{SG}(r_\mathrm{d};r_\mathrm{i})|^2$ without and with space gating, respectively, measured at the detector plane for a specific illumination point $r_\mathrm{i}$. Figure 2d1 and 2e1 are the intrinsic system PSFs through a transparent medium composed of polyacrylamide (PAA) gel. The FWHMs of the foci, which dictate the imaging resolution, were measured to be 1.5 μm either without or with space gating. In Figs. 2d2-2d4 and 2e2-2e4, we introduced an optical inhomogeneity using a scattering poly(dimethylsiloxane) (PDMS) layers on the input and output surface of the sample cuvette (see Methods for details about the sample preparation). The distance between the input/output surfaces to the object plane was about 4 mm. The optical thicknesses of the input and output layers ($L_\mathrm{i}/l_\mathrm{s}$, $L_\mathrm{d}/l_\mathrm{s}$) were (6.9, 10.6), (6.9, 12.8), and (10.6, 12.8) for Figs. 2d2-2d4 and 2e2-2e4. The ballistic wave appeared as a peak at $r_\mathrm{d}$ conjugate to $r_\mathrm{i}$, and a fluctuating background of the multiply scattered wave was spread on the detector plane. Therefore, the spot contrast, which is defined as the averaged intensity ratio of the peak to the fluctuating background, can be used to approximate the signal to noise ratios $\tau = |E_S(r_\mathrm{d};r_\mathrm{i})|^2/|E_M(r_\mathrm{d};r_\mathrm{i})|^2$ (without space gating) and

$\tau_{SG} = \left|E_S^{SG}(r_d; r_i)\right|^2 / \left|E_M^{SG}(r_d; r_i)\right|^2$ (with space gating). We used the ratios τ and $\tau_{SG}$ to quantify the imaging fidelity throughout our experiments (see Methods for further details). For instance, for the case of $(L_i/l_s, L_d/l_s)$ ~ (10.6, 12.8), τ was about 0.1 while $\tau_{SG}$ was about 30, from which we can expect only space-gated imaging properly provides a diffraction-limited resolution in this regime of scattering.

**Imaging through a scattering medium**

To demonstrate the effect of space gating in confocal imaging, we performed imaging of amplitude objects through scattering layers with various thicknesses (Fig. 3). The amplitude objects used were 2-μm gold-coated microspheres with a transmission of ~10 % at 532 nm. The illumination beam was scanned over 16.1 μm × 16.1 μm with the step size of 0.54 μm using a pair of galvanometer mirrors, resulting in 900 illumination spots. The confocal image of the object was then reconstructed by the intensity recordings at the confocal detection point $r_d = r_i$.

The reconstructed images without and with space gating are shown in Figs. 3a and 3b, respectively. The optical thicknesses of scattering layers $(L_i/l_s, L_d/l_s)$ were (0, 0), (6.9, 10.6), (6.9, 12.8), and (10.6, 12.8) (the same configuration as the PSF measurements in Figs. 2d and 2e). In a relatively weak scattering regime (as shown in the first and second columns of Fig. 3), both methods yielded the image of gold-coated microspheres although the conventional confocal imaging presented a considerable background fluctuation. When scattering became stronger (as shown in the third and fourth columns of Fig. 3), only the space-gated confocal imaging provided the resolved image of gold-coated microspheres. It is remarkable that the objects could be clearly resolved even in the highly scattering regime of $L/l_s > 23$ with the aid of space gating, where the conventional method presented only the randomly fluctuating noise dominated by the multiply scattered wave. The imaging results are in good agreement with the PSFs measured in Fig. 2d and 2e in the sense that the well-resolved image was reconstructed only when the spot contrast ($\tau$ and $\tau_{SG}$) was sufficiently large. Interestingly, for the case of ($L_i/l_s, L_d/l_s$) ~ (6.9, 12.8), the reconstructed image (Fig. 3a3) was significantly degraded even though the ballistic spot was distinctively visible (i.e. $\tau = 9.1 > 1$ as shown in Fig.2d3). This is because the multiply scattered wave $E_M(r_d; r_i)$ with a relatively small amplitude can cause the large fluctuation in $|E_S(r_d; r_i) + E_M(r_d; r_i)|^2$ depending on the relative phase of $E_M(r_d; r_i)$ to $E_S(r_d; r_i)$.

**Quantitative assessment of noise suppression factor by space gating**

To elucidate the effect of space gating depending on the degree of scattering, we estimated $\tau$ and $\tau_{SG}$ for a wide range of total optical thickness, $L_{tot}/l_s$ with $L_{tot} = L_i + L_d$. We fixed the optical thickness of the input layer $L_i/l_s$ to 6.9, 10.0, and 10.6 and varied the optical thickness of the output layer $L_d/l_s$ for each case of input layer. For the convenience of analysis, we set $L_i/l_s < L_d/l_s$ in our experiments, such that $\Delta w_{M_i} < \Delta w_{M_d}$ for all the cases. In Fig. 4a, three lines in different markers show $\tau$ and $\tau_{SG}$ for the three cases of $L_i/l_s$. In every case, $\tau_{SG}$ lies well above $\tau$, proving the effectiveness of space gating. $\tau$ was monotonically decreased with $L_{tot}/l_s$, and its behavior was generally dictated by the exponential decay of the ballistic wave because the decay of the multiply scattered wave is much slower than the exponential rate. However, $\tau_{SG}$ was highly varying depending on $L_i/l_s$. For instance, at $L_{tot}/l_s$ of 21, $\tau_{SG}$ was 36 for $L_i/l_s$=6.9 and 240 for $L_i/l_s$=10.0. This supports our theoretical prediction in Eq. (5) that the effect of space gating is mainly determined by $\Delta w_{M_i}(< \Delta w_{M_d})$, which was set by $L_i/l_s$, rather than $L_{tot}/l_s$, in our experimental configuration.

Figure 4b presents $\eta$ obtained from $\tau_{SG}/\tau$. $\eta$ ranged from 4.4 to 150 depending on the combination of the input and output scattering layers. As predicted in Eq. (5), $\eta$ was largely determined by the $L_i/l_s$, and $L_d/l_s$ had a marginal impact on $\eta$ as $L_i/l_s < L_d/l_s$. At $L_i/l_s$ of 6.9, $\eta$ was in the range of 4~11, while it was increased to 47~100 when $L_i/l_s$ was increased to 10 or 10.6. Similar to $\tau_{SG}$, even for the same $L_{tot}/l_s$ (e.g. at $L_{tot}/l_s$ of 21 in our experiments), $\eta$ can be varied significantly, implying that the effect of space gating is highly dependent on the axial position of the object plane within a homogeneously scattering medium. The maximum noise suppression factor we observed was $\eta = 150$ for the configuration of $(L_i/l_s, L_d/l_s) = (10.6, 13.9)$.

The quantitative behavior of $\tau$, $\tau_{SG}$, and $\eta$ highly depends on the optical properties of sample, and the configuration of sample and imaging system. On the other hand, once the transfer functions of scattering medium, $T_i(r_o; r_i)$ and $T_d(r_o; r_d)$, are determined, one may precisely estimate $\tau$, $\tau_{SG}$, and $\eta$ using the model presented in the Principle section.

**Imaging of amplitude and phase objects *embedded* inside a scattering medium**

For most of practical applications, objects are completely embedded within a scattering medium. In such cases, the speckle grain on the object plane becomes as small as half the wavelength of the probing laser, which has posed difficulties in the previous optical scale acousto-optic wavefront shaping and imaging modalities. To verify that our imaging scheme is robust against the small speckle grains inside an acoustic focus, we have further demonstrated imaging of gold-coated microspheres fully embedded within a scattering medium (as shown in Fig. 5a). A thin PAA gel layer mixed with gold beads was sandwiched between the two 3 mm-thick PAA gel slabs containing 0.8% of fat emersion (Intralipid). The total optical thickness $L_{tot}/l_s$ of the 6-mm-thick PAA gel (shown in the bottom right inset of Fig. 5a) was measured to be 21.0. We recorded the speckle pattern at the object plane with a 1.4-NA objective lens in the absence of the PAA slab on the detection side (bottom left inset of Fig. 5a) and determined the average grain size as the FWHM of autocorrelation function of the speckle pattern. We confirmed that the speckle grain was 280 nm in width, which is about half the wavelength. While the image of the microspheres was completely scrambled by multiple scattering without space gating (Fig. 5b), they were clearly resolved with space gating (Fig. 5c). $\tau_{SG}$ and $\tau$ were estimated to be 260 and 1.1 from the measured PSFs, leading to the noise suppression factor $\eta$ of 240.

By leveraging the noise suppression capability of space gating and the coherent treatment of ballistic wave, we demonstrate the unique capability of space gating – quantitative phase imaging of objects completely embedded within a scattering medium. We prepared human red blood cells sandwiched between the PAA gels with 0.8% of fat emersion ($L_{tot}/l_s \sim 21$) to mimic biological conditions. As shown in Fig. 5c, only the speckled phase pattern was visible without space gating due to the dominance of intense multiply scattered wave over the ballistic wave. On the contrary, our method revealed the phase delay, and in turn, the morphology of the red blood cells embedded within the scattering medium (Fig. 5d). To our knowledge, this is the first experimental demonstration of the quantitative phase imaging of single biological cells embedded within such a thick scattering medium. This opens a new venue for interrogating transparent biological cells within small animals or organs with no use of exogenous contrast agents.

**Discussion**

In our study, we proposed and implemented the concept of space gating using a high frequency acoustic focus. The space gating is distinct from the existing gating operations in that the gating is applied directly at the object plane rather than the detection plane. Therefore, one may use space gating to filter out the multiply scattered wave that bypasses the existing gating methods. We showed that the space gating reduced the relative intensity of multiply scattered wave to the ballistic wave by more than 100 times and thus enabled diffraction-limited imaging of the objects fully embedded within a scattering medium of $L_{tot}/l_s$ of 21.0. Furthermore, combining the advantages of the effective rejection of multiply scattered wave and the coherent treatment of ballistic wave, we demonstrated an unprecedented capability that would serve as a unique tool in the field of bioimaging – imaging biological phase objects embedded within a scattering medium.

The proposed space gating method is the first acousto-optic imaging approach that relies on ballistic wave. Its resolution is dictated by the ideal diffraction limit of the optical system, not by the acoustic diffraction limit. However, it is worth noting that the conventional acousto-optic and photoacoustic approaches, relying on both ballistic and multiply-scattered wave as a whole, are not subject to the problem of competition between the ballistic and multiply scattered waves. Therefore, their imaging depth is larger than the proposed method, albeit at the expense of resolving power. There have been a few ingenious wavefront shaping methods that can improve the spatial resolution of acousto-optic or photoacoustic approaches to the optical speckle scale using iterative optimization[32–34] and variance-encoding[35,36]. However, these methods can easily be compromised in practical situations where the size of speckle grain is as small as the optical wavelength or the acoustic focal profile does not have a well-defined peak. Those concepts have only been demonstrated in the geometry where the gap between the scattering layer and the object plane is large enough so that the speckle grains are at least one order of magnitude larger than the wavelength[33–37]. On the contrary, our method, relying on a ballistic wave for image formation, enables us to obtain the ideal optical diffraction-limited resolution for the objects completely embedded within a scattering medium, where the speckle grains are fully developed and their average size is close to half the wavelength. Furthermore, our method can be much less sensitive to speckle decorrelation compared to the acousto-optic wavefront manipulation techniques because the dynamic motion of the scatters affects much less to the ballistic wave than the multiply scattered wave.

The space gating can be far more effective than the coherence gating, or equivalently the time gating, in the case of a forward scattering sample such as biological tissues. Similar to the relation in Eq. (5) of the space gating, the effect of coherence gating can be estimated as the ratio $\Delta t_M/\Delta t_{SG}$, where $\Delta t_M$ and $\Delta t_{SG}$ are respectively the spread of multiply scattered wave and the width of gating window in the time domain. Based on geometric optics, the scattered ray with a deflection angle $\theta$ experiences the temporal spread of $\Delta t_M \sim L(1/\cos\theta - 1)/c$, where c is the average speed of light in the medium. For a small $\theta$, $\Delta t_M$ is approximated as $L\theta^2/2$. On the other hand, the spatial spread $\Delta w_M$ in the transverse plane of object is given as $L/2 \times \tan\theta$, which is approximately $L\theta/2$ for a small $\theta$. Plus, the space gating is a two-dimensional operation on the object plane while the time gating is one-dimensional in the time domain (i.e. the effect of space gating is given as the square of $\Delta w_M/\Delta w_{SG}$). Assuming that the gating windows for coherence gating and space gating have comparable sizes (typically, in tens of micrometers), the space gating can increasingly outperform the time gating as the medium becomes more forward scattering. To our estimation, the noise suppression factor of the 30 μm-width coherence gating is only around 30 for the case of scattering sample used in Fig. 5, while it was measured to be 240 when the space gating with the window size of ~ 30 μm × 70 μm was used. The effectiveness of space gating in a forward scattering regime is one of the main reasons why the phase imaging within a scattering medium, which could have not been realized with the coherence gating, was enabled with the space gating.

Because of the two-dimensional nature of space gating, the noise suppression factor η can be quadratically improved by reducing the size of gating window $\Delta w_{SG}$. Therefore, the use of higher frequency acoustic wave or second-harmonic acousto-optic interaction would greatly improve the imaging depth. The imaging depth can also be greatly improved by choosing the longer wavelength of the probing beam at which $l_s$ is larger. First, it allows us to detect the ballistic wave at a proportionally larger $L$ because the intensity of the ballistic wave follows the Beer-Lambert law dictated by $L/l_s$. Secondly, and more interestingly, the effect of space gating would be quadratically increased with $L$ due to the associated increase in the spatial extent $\Delta w_M$ of the multiply scattered wave. Therefore, although our proof-of-concept experiments was performed at the wavelength of 532 nm where $l_s$ is relatively small for biological tissues, the use of longer-wavelength source would be much more beneficial in increasing the absolute imaging depth in biological applications. The space gating could also be adopted for the epi-detection configuration for more diverse applications in biological studies.

The resolution of the demonstrated imaging method is set by the diffraction limit of optical system. In the present study, the diffraction-limited resolution of 1.5 μm was set by the geometric restriction of the focused laser beam and the acoustic transducer. The use of an acoustic transducer of a smaller physical size will allow the higher numerical aperture for optical imaging. Novel aberration correction methods reported in the previous studies can also be incorporated to retain a sub-micron imaging resolution even for aberrating biological specimens [18]. In our experiments, the image acquisition time for each object point ranged between 100 ms to 1 s depending on the exposure time set by the optical thickness of the scattering medium, but it can be improved using an off-axis holographic measurement or a high-repetition laser sources.

To conclude, the imaging depth of microscopy has long been set by the ability of exiting gating methods to reject a multiply scattered wave. Especially, it has been even more difficult to perform phase imaging of transparent biological cells inside a scattering medium due to its susceptibility to multiple scattering. The proposed concept of space gating is a novel and independent gating scheme, which can effectively reject the multiply scattered wave that bypasses the conventional gating operations, and in turn, enables the unprecedented imaging capability of phase imaging within a scattering medium. With the potential to be combined with the existing gating methods, the future development and use of space gating will provide an important step towards reaching the ultimate imaging depth set by the detection limit of ballistic wave and promote the applicability of phase imaging for studying the native physiology of the biological cells within deep tissues.

## Methods

### Measurement of transfer functions

To measure the illumination transfer function $|T_i(r_o; r_i)|^2$, we recorded the intensity map on the object plane using the camera shown in Fig. 2 while the scattering sample on the detection path was removed. To measure the detection transfer function $|T_d(r_o; r_d)|^2$, we used the reciprocity of light propagation and the symmetry in our optical system. Based on the reciprocity, the detection transfer function $|T_d(r_o; r_d)|^2$ is identical to the intensity map on the object plane for a virtual source placed at the detector point $r_d$. Therefore, we removed the scattering sample on the illumination path and flipped the entire sample with respect to the object plane to use the symmetry in the input and output side of our system. Finally, similar to the measurement of $|T_i(r_o; r_i)|^2$, we recorded the intensity map on the object plane while illuminating the flipped sample with a focused beam.

### Calculation of modulation efficiency

The measured interference intensity at $k^{th}$ phase step (k is an integer number $\in [0,3]$) can be expressed as $I_k = \left|E^{ref}\exp\left(i\frac{\pi}{2}k\right) + E^{sam}\right|^2 = \left|E^{ref}\exp\left(i\frac{\pi}{2}k\right) + E^{sam}_{unmod} + E^{sam}_{mod}\right|^2$, where $E^{ref}$ and $E^{sam}$ are the complex amplitudes of reference and sample waves, respectively, and $E^{sam}_{unmod}$ and $E^{sam}_{mod}$ are respectively the unmodulated and modulated components of the sample wave. Then, the modulation efficiency is defined as $|E^{sam}_{mod}|^2/|E^{sam}|^2$. Considering the camera exposure is much longer than the acoustic oscillation period, the two interference terms involving $E^{sam}_{unmod}$ are averaged out to the negligible level due to their oscillation at the acoustic frequency. Therefore, $I_k$ can be written as $|E^{ref}|^2 + |E^{sam}|^2 + 2|E^{ref}||E^{sam}_{mod}|\cos(\phi + \frac{\pi}{2}k)$, where $\phi$ is the relative phase between $E^{\text{ref}}$ and $E^{\text{sam}}_{\text{mod}}$. Finally, the modulation efficiency is given by $\left\{\frac{[(I_2-I_0)+i(I_3-I_1)]}{4}\right\}^2 / |E^{\text{ref}}|^2 |E^{\text{sam}}|^2$.

**Scattering layer preparation**

To fabricate a scattering layer, poly(dimethylsiloxane)(PDMS) solution were thoroughly mixed with ZnO particles at a fixed concentration. The mixture was transferred to a Petri dish and coated uniformly on the dish using a spin coater. Finally, PDMS was cured at 80°C. The scattering mean free path of the layer was $l_s = 21$ μm, which was measured by the ballistic transmission through two distant diaphragms. The layer thickness was controlled by varying the volume of PDMS mixture transferred to the dish. The thickness was then measured using a conventional bench-top microscope, and it ranged between 150 μm and 290 μm.

**Calculation of ratio τ and $\tau_{\text{SG}}$ using the PSFs**

The detected confocal intensity $\langle |E(r_d = r_i; r_i)|^2 \rangle$ equals to $\langle |E_S(r_d = r_i; r_i)|^2 \rangle + \langle |E_M(r_d = r_i; r_i)|^2 \rangle$ because the cross-term between the ballistic and multiply scattered wave converges to 0 with an ensemble average denoted by $\langle\ \rangle$. $\langle |E_M(r_d = r_i; r_i)|^2 \rangle$ here can be separately determined by the intensity in the vicinity ($r_d \sim r_i$) of the illumination spot because the $\langle |E_M(r_d; r_i)|^2 \rangle$ is slowly varying with $r_d$.

The ratio of ballistic to multiply scattered waves was calculated using two methods depending on the visibility of the focused ballistic wave. When the focused spot was clearly visible (i.e. the peak to background ratio was higher than 2), the detected confocal intensity $\langle |E(r_d = r_i; r_i)|^2 \rangle$ and the $\langle |E_M(r_d \sim r_i; r_i)|^2 \rangle$ can be quantified directly from the PSFs. Therefore, $\tau$ and $\tau_{\text{SG}}$ are determined as $[\langle |E(r_d = r_i; r_i)|^2 \rangle - \langle |E_M(r_d \sim r_i; r_i)|^2 \rangle] / \langle |E_M(r_d \sim r_i; r_i)|^2 \rangle$ (for $\tau_{\text{SG}}$, $[\langle |E^{\text{SG}}(r_d = r_i; r_i)|^2 \rangle - \langle |E^{\text{SG}}_M(r_d \sim r_i; r_i)|^2 \rangle] / \langle |E^{\text{SG}}_M(r_d \sim r_i; r_i)|^2 \rangle$). However, when the focused

spot was not well visible such as in Fig.2d4, $\langle |E_S(r_d = r_i; r_i)|^2 \rangle$ cannot be precisely estimated by $\langle |E(r_d = r_i; r_i)|^2 \rangle - \langle |E_M(r_d \sim r_i; r_i)|^2 \rangle$. In this case, $\langle |E_S(r_d = r_i; r_i)|^2 \rangle$ was estimated by $I_0 \exp(-L_{tot}/l_s)$, where $I_0$ is the measured peak intensity through a transparent specimen. Then, $\tau$ was estimated as $I_0 \exp(-L_{tot}/l_s)/\langle |E_M(r_d \sim r_i; r_i)|^2 \rangle$.

**Acknowledgments**

This research was supported by IBS-R023-D1. M.J. was supported by TJ Park Science Fellowship of POSCO TJ Park Foundation.

**Author contributions**

M.J. and W.C. conceived the initial idea. M.J. developed theoretical modeling, designed the experiments, and analyzed the experimental data with the help of H.K. and W.C. H.K. prepared the sample and carried out the experiments with the help of M.J. W.K.L. and J.L. fabricated the gold beads. All authors contributed to writing the manuscript. W.C. supervised the project.

**Competing financial interests**

The authors declare no competing financial interests.

**Figures**

**Figure1**

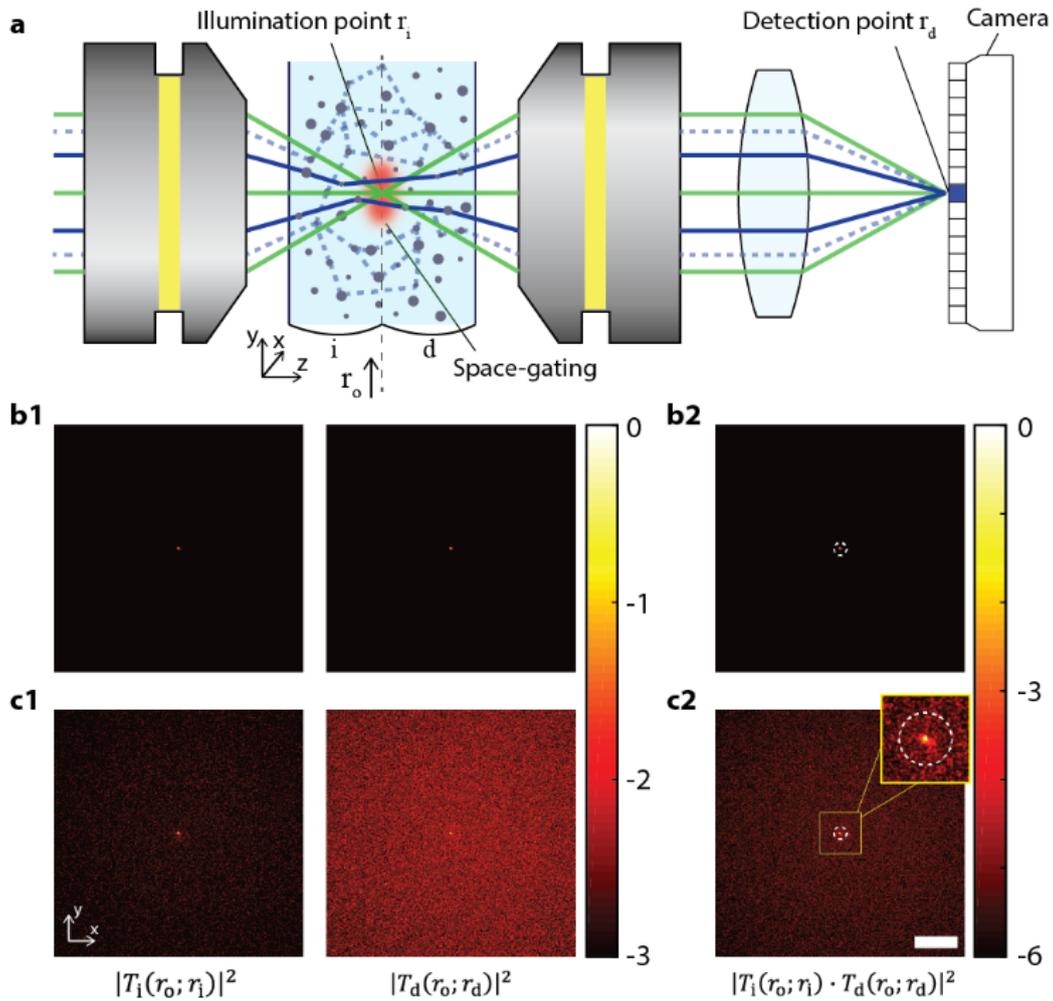

**Fig. 1 | Principle of space gating. a,** Schematic of the imaging principle. Conventional confocal imaging methods relies on the ballistic waves (shown as green lines). When optical inhomogeneity is introduced, the intensity of the ballistic wave is exponentially decreased with depth, and the multiply scattered wave (shown as solid blue and dotted blue lines) may obscure the ballistic wave. By implementing space gating at the object plane using an acousto-optic effect (indicated as a red spot), the multiply scattered wave that travels outside the acoustic focus (dotted blue lines) is

rejected, which in turn improves the ratio of the ballistic wave to the multiply scattered wave at the sensor element (marked as blue pixel) whose position is conjugate to the illumination point $r_d \sim r_i$. **b1,** Intensity maps of illumination and detection transfer functions in a confocal gating scheme (where $r_d \sim r_i$) with respect to $r_o$ on the object plane for a transparent medium. **b2,** Contribution map, $|T_i(r_o; r_i)T_d(r_o; r_d)|^2$, with respect to $r_o$ calculated from the transfer functions in **b1**. **c1-c2,** Same as **b1** and **b2**, respectively, but in the presence of scattering. The optical thicknesses on the illumination and detection sides were $L_i/l_s = 10.6$ and $L_d/l_s = 12.8$, respectively. Scale bar, 100 μm.

**Figure2**

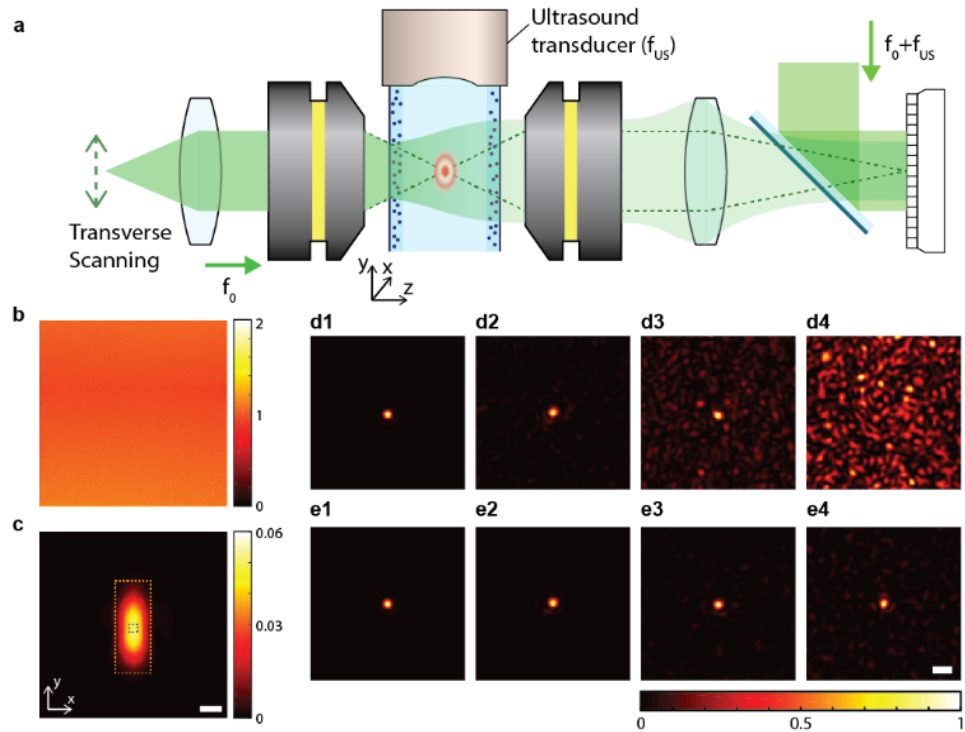

**Fig. 2 | Coherent space-gated microscopy and its point spread functions. a,** Focused acoustic beam modulates the frequency of the incident focused illumination beam whose optical frequency is $f_0$. Only the frequency-modulated wave through the region of the space gating (i.e. acoustic focus) are measured at the sensor plane by using a phase-shifting interferometry, where the frequency of the reference beam is set to that of the acoustically modulated optical wave $f_0 + f_{us}$. **b,** Average intensity map for 900 planar illuminations with different incidence angles through a transparent medium without space gating. Without space gating, the entire object plane contributes to the detected signal. The intensity map is normalized to the mean intensity. **c,** Same as **b**, with space gating. With the space gating, only the region inside the gating window (i.e. acoustic focus) contributes to the detected signal. The intensity map was normalized such that the map represents the acoustic modulation efficiency. Scale bar: 30 μm **d1-d4,** Point spread functions $|E(r_d; r_i)|^2$

measured on the detection plane without space gating, when the optical thicknesses of the input and output layers were (0, 0), (6.9, 10.6), (6.9, 12.8), and (10.6, 12.8), respectively. **e1-e4,** Point spread functions $|E_{SG}(r_d; r_i)|^2$ with space gating for the scattering layers corresponding to **d1-d4**. Point spread functions were normalized to their maximum intensities. Scale bar: 5 μm.

Figure3

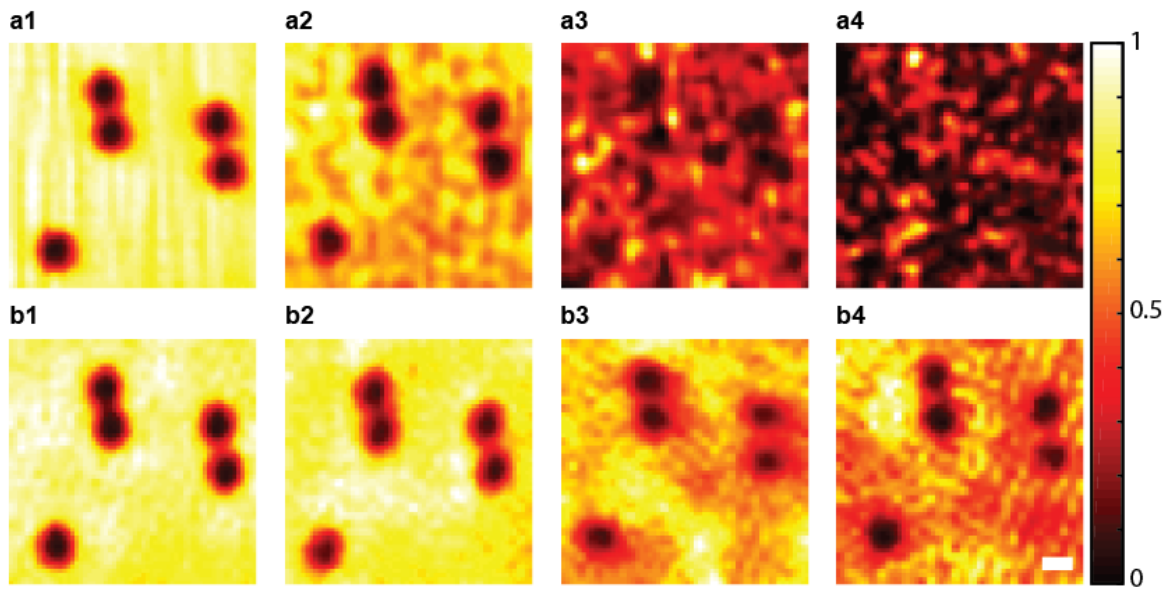

**Fig. 3 | Demonstration of the effect of space gating in confocal imaging.** Images were reconstructed by scanning 900 points within the field of view of 16.1 μm × 16.1 μm. **a1-a4**, Reconstructed intensity images of 2-μm gold-coated microspheres without space gating when the optical thicknesses of the input and output layers were (0, 0), (6.9, 10.6), (6.9, 12.8), and (10.6, 12.8), respectively. **b1-b4**, Reconstructed images with space gating at the scattering configurations same as in **a1-a4**. Images were normalized to their maximum intensities. Scale bar: 2 μm.

**Figure 4**

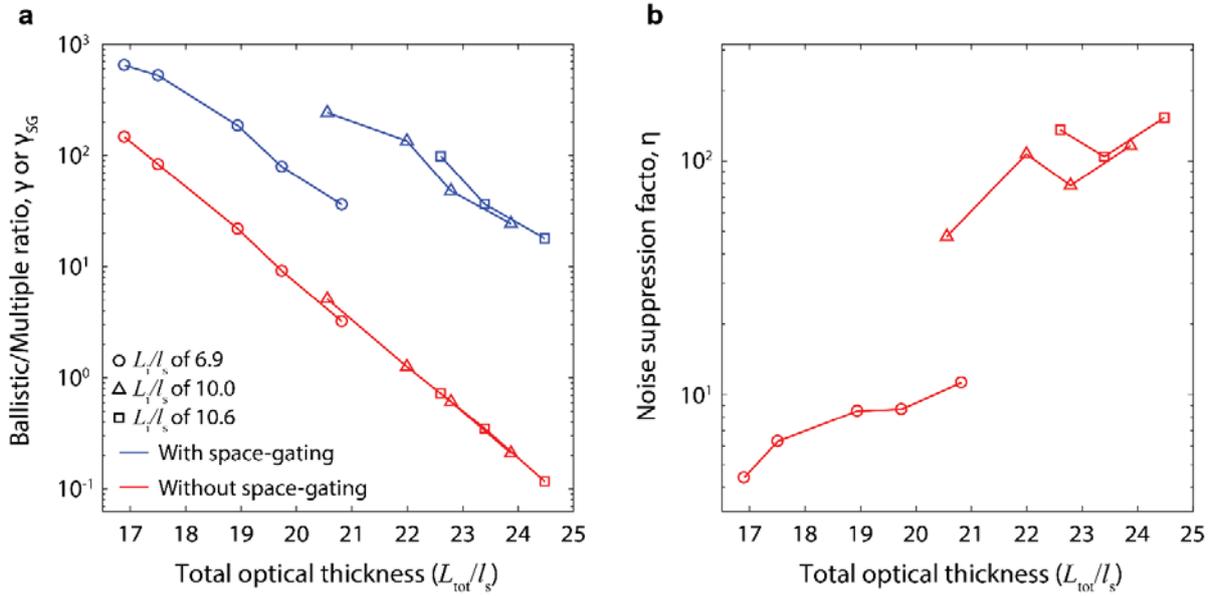

**Fig. 4 | Dependence of the effect of space gating on the optical thickness of scattering medium. a,** Ratio of the ballistic to multiply scattered waves with ($\tau_{SG}$, blue) and without space gating ($\tau$, red) as a function of the total optical thickness, $L_{tot}/l_s$. Circular, triangular and rectangular markers indicate the cases when the optical thicknesses of the input layer, $L_i/l_s$, were fixed to 6.9, 10.0, and 10.6, respectively. The optical thickness of the output layer was varied for each case. **b,** Noise suppression factor $\eta$ of space gating. $\eta$ was obtained from $\tau_{SG}/\tau$ in **a**.

Figure5

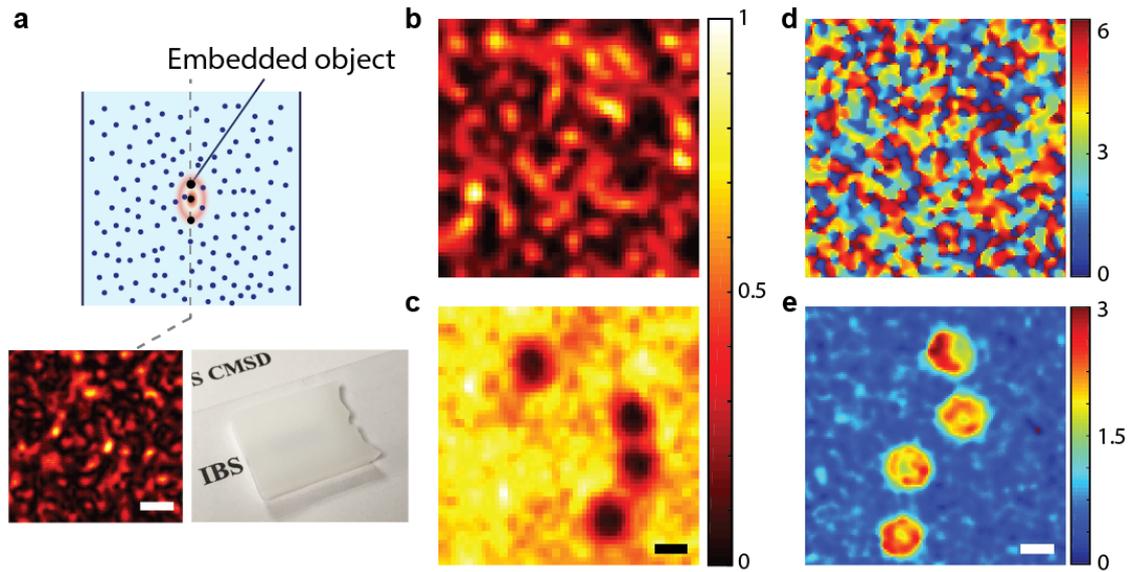

**Fig. 5 | Demonstration of coherent space-gated imaging of objects fully embedded within a scattering medium. a,** Schematic of the sample configuration. The bottom left inset shows the speckle pattern measured right at the object plane after removing the right-hand side of the scattering medium (scale bar in the inset, 1 μm), and the bottom right inset shows the photograph of the scattering medium. The optical thickness of the scattering slab was 21.0. **b** and **c,** Reconstructed images of 2-μm gold-coated microspheres embedded within the scattering medium without and with space gating, respectively. With the noise suppression factor $\eta=240$ of space gating, the gold-coated beads were clearly resolved. Images were normalized to their maximum intensities. Scale bar: 2 μm. **d** and **e,** Reconstructed phase images of human red blood cells embedded within the same scattering medium used in **b** and **c,** without and with space gating, respectively. The size and the morphology of the red blood cells can be obtained from the phase map. Scale bar: 5 μm.